\documentclass[11pt]{article}
\usepackage{graphicx}

0.1cm \evensidemargin - 5cm
\newcommand{\eproof}{\rule{0.2cm}{0.2cm}}

\newtheorem{thm}{Theorem}[section]
\newtheorem{prop}[thm]{Proposition}

\newtheorem{cor}[thm]{Corollary}

\newtheorem{remark}[thm]{Remark}

\begin{document}

\title{\Large{{\bf On a representation of the inverse $F_{q}$-transform}}}
\author{Sabir Umarov$^{1}$, Constantino Tsallis$^{2,3}$}
\date{}
\maketitle
\begin{center}
$^{1}$ {\it Department of Mathematics,
Tufts University, Medford, MA 02155, USA\\}
$^{2}$ {\it Centro Brasileiro de Pesquisas Fisicas,
Xavier Sigaud 150, 22290-180 Rio de Janeiro-RJ, Brazil}\\
$^{3}$ {\it Santa Fe Institute,
1399 Hyde Park Road, Santa Fe, NM 87501, USA}
\end{center}

\begin{abstract}
A representation formula for the inverse $q$-Fourier
transform is obtained in the class of functions
$\mathcal{G}=\bigcup_{1\le q<3}\mathcal{G}_q,$ where
$\mathcal{G}_{q}=\{f = a e_{q}^{-\beta x^2}, \, a>0, \, \beta>0 \}.$
\end{abstract}

\vspace{1cm}

In this paper we find a representation formula for the inverse
$F_q$-transform in a class of functions $\mathcal{G}$ defined below.
$F_q$-transform is a useful tool in the study of limit processes in nonextensive statistical mechanics (see \cite{nonextensive} and references therein). Note that, in this
theory, random states are correlated in a special manner, and
a knowledge on $F_q$-inverse of data is helpful in understanding
the nature of such correlations.

Throughout the paper we assume that $1 \leq q < 3.$ The {\it
$F_q$-transform}, called also {\it $q$-Fourier transform}, of a
nonnegative $f(x) \in L_1(R^1)$ is defined by the formula (see
\cite{UmarovTsallisSteinberg2006,UmarovTsallis2007})
\begin{equation}
\label{Fourier} F_q[f](\xi) = \int_{supp \, f} e_q^{ix\xi} \otimes_q
f(x) dx \,,
\end{equation}
where $\otimes_q$ is the symbol of the $q$-product, and $e_q^{x}$ is
a $q$-exponential. The reader is referred for details of $q$-algebra
and $q$-functions to \cite{nivanen,borges,Tsallis2005}. $F_q$ coincides with
the classic Fourier transform if $q=1.$ If $1<q<3$ then $F_q$ is a
nonlinear mapping in $L_1.$ The representation formula for the
inverse, $F_q^{-1}$, is defined in the class of function of the form
$A e_{q_1}^{-B \xi ^2},$ since it uses a specific operator $I$
defined in this class. The question on extension of this operator to
wider classes is remaining a challenging question.

The obvious equality $e_q^{ix\xi} \otimes_q f(x) = f(x)
e_q^{ix \xi[f(x)]^{q-1}}$, which holds for all $x \in supp
\, f,$ implies the following lemma, which gives an expression for
the $q$-Fourier transform without usage of the $q$-product.
\begin{prop}
\label{informal} The $q$-Fourier transform can be written in the
form
\begin{equation}
\label{identity2} F_q[f](\xi) = \int_{supp \,f} f(x) e_q^{ix
\xi[f(x)]^{q-1}} dx.
\end{equation}
\end{prop}
\vspace{.2cm}

Introduce the operator
\begin{equation}
\label{identity2} F_q^{\ast}[f](\xi) = \int_{supp \,f} f(x) e_q^{-ix
\xi[f(x)]^{q-1}} dx.
\end{equation}

For arbitrary nonnegative $f \in L_1 (R)$ both operators, $F_q$ and
$F_q^{\ast}$ are correctly defined. Moreover,
\begin{equation}
\sup_{\xi \in R^1} |F_q[f](\xi)| \leq \|f\|_{L_1} \, \, \,
\mbox{and} \, \, \, \sup_{\xi \in R^1} |F_q^{\ast}[f](\xi)| \leq
\|f\|_{L_1}.
\end{equation}

Introduce the set of functions
\begin{equation}
\mathcal{G}_q =\{f: f(x)=a e_q^{-\beta x^2}, \, a>0, \, \beta>0\}.
\end{equation}

Obviously, $\mathcal{G}_q \subset L_1$ for all $q<3$. The set
$\mathcal{G}_q$ is fully identified by the triplet $(q,a,\beta).$ We
denote
$$\mathcal{R} = \{(q,a,\beta): q < 3, a >0, \beta >0 \}.$$ For any function $f \in \mathcal{G}_q$ we
have $f(-x)=f(x)$, so $f$ is symmetric about the origin. Moreover,
it follows from the symmetry that $F_q[f](\xi) =
F_q^{\ast}[f](\xi).$ Further, a function $f \in \mathcal{G}_q$ with
$a=\frac{\sqrt{\beta}}{C_q}$, where
\begin{equation}
C_q =
\left\{\begin{array}{ll}
   {
\frac{2}{\sqrt{1-q}} \int_0^{\pi / 2} (\cos \, t)^{\frac{3-q}{1-q}} dt }
= \frac{2\sqrt{\pi}\, \Gamma\bigl({1 \over {1-q}}\bigr)}
{(3-q) \sqrt{1-q} \, \Gamma\bigl({{3-q} \over {2(1-q)}}\bigr)}, & -\infty<q<1, \\

{\sqrt{\pi}}, & q=1, \\

\frac{2}{\sqrt{q-1}} \int_0^{\infty} (1+y^2)^{{-1} \over {q-1}} dy =
\frac{\sqrt{\pi} \, \Gamma\bigl(\frac{3-q}{2(q-1)}\bigr)}{\sqrt{q-1}
\, \Gamma \bigl({1 \over {q-1}}\bigr)}
, &  1<q<3 \,. \\
\end{array} \right.
\end{equation}
is called a $q$-Gaussian, and is denoted by $G_q(\beta;x)$. Thus, the set
of all $q$-Gaussians forms a subset of $\mathcal{G}_q.$

The following statement was proved in
\cite{UmarovTsallisSteinberg2006}.

\begin{prop}
\label{centrallemma} Let $1 \leq q < 3.$ For the $q$-Fourier
transform of a $q$-Gaussian, the following formula holds:
\begin{equation}
\label{gausstransform0}
F_q[G_q(\beta; x)](\xi) = \Bigl(e_q^{\, -\frac{\xi^2}{4 \beta^{2-q}
C_q^{2(q-1)}}}\Bigr)^{{3-q \over 2}}.
\end{equation}
\end{prop}

Assume a sequence $q_k$ is given by
\begin{equation}
\label{qk}
q_k = \frac{2q-k(q-1)}{2-k(q-1)}, \, -\infty < k <
\frac{2}{q-1}-1,
\end{equation}
for $q>1$, and $q_k = 1$ for all $ k=0, \pm 1,...,$ if $q=1.$

It follows from this proposition the following result.

\begin{cor}
\label{centrallemmacor} Let $1 \leq q <3$ and $k < \frac{2}{q-1}-1.$
Then
\begin{equation}
\label{gausstransform21} F_{q_k}[G_{q_k}(\beta; x)](\xi) =
e_{q_{k+1}}^{- \beta_{k+1} \xi^2},
\end{equation}
where  $q_{k+1}=\frac{1+q_k}{3-q_k}$ and $\beta_{k+1} =
\frac{3-q_k}{8 \beta^{2-q_k} C_{q_k}^{2(q_k-1)}}.$
\end{cor}

\begin{remark}
It follows from (\ref{gausstransform21}) that
\[
F_{q_k}[a e_{q_k}^{-\beta x^2}](\xi) = \frac{aC_{q_k}}{\sqrt{\beta}}
\, \, \,  e_{q_{k+1}}^{- B \xi^2},
\]
where $B=\frac{a^{2(q_k-1)}(3-q_k)}{8 \beta}.$

\end{remark}

\begin{thm}
The operator $F_{q_k}:\mathcal{G}_{q_k} \rightarrow
\mathcal{G}_{q_{k+1}}$ is invertible.
\end{thm}

{\em Proof.} With the operator $F_{q_k}: \mathcal{G}_{q_k}
\rightarrow \mathcal{G}_{q_{k+1}}$ we associate the mapping
$\mathcal{R} \rightarrow \mathcal{R}$ defined as $(q_k,a,\beta)
\rightarrow (q_{k+1},A,B),$ where $A=\frac{aC_{q_k}}{\sqrt{\beta}}$
and $B=\frac{a^{2(q_k-1)}(3-q_k)}{8 \beta}.$
Consider the system of equations
\[
\frac{1+q_k}{3-q_k}=Q,
\]
\[ \frac{aC_{q_k}}{\sqrt{\beta}}=A,
\]
\[
\frac{a^{2(q_k-1)}(3-q_k)}{8 \beta}=B
\]
with respect to $q_k, a, \beta$ assuming that $Q, A$ and $B$ are
given. The first equation is autonomous and has a unique solution
$q_k=(3Q-1)/(Q+1).$ If the condition $k < \frac{2}{q-1}-1$ is
fulfilled then the other two equations have a unique solution as
well, namely
\begin{equation}
\label{abeta}
a=\left(\frac{A\sqrt{3-q_k}}{2C_{q_k}\sqrt{2B}}\right)^{\frac{1}{2-q_k}},
\,\,
\beta=\left(\frac{A^{2(q_k-1)}(3-q_k)}{8C_{q_k}^{2(q_k-1)}B}\right)^{\frac{1}{2-q_k}}.
\end{equation}
It follows from (\ref{qk}) that $Q$ and $q_k$ are related as
$Q=q_{k+1}.$ Hence, the inverse mapping $(F_q)^{-1}:
\mathcal{G}_{k+1} \rightarrow \mathcal{G}_k$ exists and maps each
element $A e_{q_{k+1}}^{-B\xi^2} \in \mathcal{G}_{k+1}$ to the
element $a e_{q_k}^{-\beta x^2} \in \mathcal{G}_k$ with $a$ and
$\beta$ defined in (\ref{abeta}). \eproof

 Now we find a representation formula for the inverse
operator $F^{-1}_q.$ Denote by $T$ the mapping $T:(a,\beta)
\rightarrow (A,B),$ where $A=\frac{aC_{q_k}}{\sqrt{\beta}}$ and
$B=\frac{a^{2(q_k-1)}(3-q_k)}{8 \beta}$, as indicated above. We have seen that $T$ is
invertible and $T^{-1}: (A,B) \rightarrow (a,\beta)$ with $a$ and
$b$ in (\ref{abeta}). Assume $(\bar{a}, \bar{\beta})= T^{-2}(A,B) =
T^{-1}(T^{-1}(A,B))=T^{-1}(a,\beta).$ Further, we introduce the
operator $I_{(q_{k+1},{q_{k-1}})}: \mathcal{G}_{q_{k+1}} \rightarrow
\mathcal{G}_{q_{k-1}}$ defined by the formula
\begin{equation}
\label{i} I_{(q_{k+1},{q_{k-1}})}[A e_{q_{k+1}}^{-B \xi^2}] =
\bar{a} e_{q_{k-1}}^{- \bar{\beta} \xi^2}.
\end{equation}

Consider the composition $H_{q_k} = F_{q_{k-1}}^{\ast} \circ
I_{(q_{k+1},{q_{k-1}})}.$ By definition, it is clear that
$I_{(q_{k+1},{q_{k-1}})}:\mathcal{G}_{q_{k+1}} \rightarrow
\mathcal{G}_{q_{k-1}}.$ Since
$F_{q_{k-1}}^{\ast}:\mathcal{G}_{q_{k-1}} \rightarrow
\mathcal{G}_{q_{k}}$, we have $H_{q_{k}}:\mathcal{G}_{q_{k+1}}
\rightarrow \mathcal{G}_{q_{k}}.$ Let $\hat{f} \in
\mathcal{G}_{k+1}$, that is $\hat{f}(\xi)=A e_{q_{k+1}}^{-B \xi^2}.$
Then, taking into account the fact that $supp \, \, \hat{f} = R^1$
if $q \geq 1,$ one obtains an explicit form of the operator
$H_{q_k}:$
\begin{equation}
\label{explicit} H_{q_k}[\hat{f}(\xi)] (x) = \int_{-\infty}^{\infty}
\left(\bar{a} e_{q_{k-1}}^{-\bar{\beta} \xi^2}\right)
\otimes_{q_{k-1}} e_{q_{k-1}}^{-i x \xi} d\xi =
\int_{-\infty}^{\infty} I_{(q_{k+1},{q_{k-1}})}[\hat{f}(\xi)]
\otimes_{q_{k-1}} e_{q_{k-1}}^{-i x \xi} d\xi.
\end{equation}

\begin{thm}
\begin{enumerate}
\item
Let $f \in \mathcal{G}_{q_k}$. Then $H_{q_{k}} \circ F_{q_k} [f]=f;$
\item
Let $f \in \mathcal{G}_{q_{k+1}}$. Then $F_{q_{k}} \circ H_{q_{k}}
[f]=f.$
\end{enumerate}
\end{thm}

\vspace{.2cm}

{\it Proof.} 1. We need to show validity of the equation
$F_{q_{k-1}} \circ I_{(q_{k+1},q_{k-1})} \circ F_{q_k} =
\mathcal{J},$ where $\mathcal{J}$ is the identity operator in
$\mathcal{G}_{q_k}.$ This equation is equivalent to $T \circ T^{-2}
\circ T = J$ ($J$ is the identity operator in $\mathcal{R}$ with
fixed $q$), which is correct by construction.

2. Now the equation $F_{q_k} \circ F_{q_{k-1}} \circ
I_{(q_{k+1},q_{k-1})}
 = \mathcal{J}^{'}$ ($\mathcal{J}^{'}$ is the identity operator in
 $\mathcal{G}_{q_{k+1}}$) is equivalent to the identity $T^2 \circ T^{-2} = J.$
 \eproof

\begin{cor}
The operator $H_{q_{k}}: G_{q_{k+1}} \rightarrow G_{q_k}$ is the
inverse to the $q_{k}-$Fourier transform: $H_{q_{k}}=F_{q_k}^{-1}.$
\end{cor}

\begin{cor}
For $q=1$ the inverse $F_{q_k}^{-1}$ coincides with the classical
inverse Fourier transform.
\end{cor}

{\it Proof.} If $q=1$, then by definition one has
$q_{k}=q_{k-1}=q_{k+1}=1.$ We find $\bar{a}$ and $\bar{\beta}$
taking $(A,B)=(1,1).$ It follows from relationships (\ref{abeta})
that $(a,\beta)= T^{-1}(1,1)=(\frac{1}{2 \sqrt{\pi}}, \frac{1}{4}).$
Again using (\ref{abeta}) we obtain
$(\bar{a},\bar{\beta})=T^{-2}(1,1)=T^{-1}(\frac{1}{2
\sqrt{\pi}},\frac{1}{4})=(\frac{1}{2\pi},1).$ This means that
$I_{(1,1)}\hat{f}(\xi)=\frac{1}{2\pi} \hat{f}(\xi).$ Hence, the
formula (\ref{explicit}) takes the form
\[
F_{1}^{-1} [\hat{f}(\xi)](x)=\frac{1}{2\pi}\int_{-\infty}^{\infty}
\hat{f}(\xi) e^{-ix\xi} dx,
\]
recovering the classic formula for the inverse Fourier transform.
\eproof

Summarizing, we have proved that, if $\hat{f}(\xi)$ is a function in
$\mathcal{G}_{q_{k+1}},$ where $q_k$ with $k < \frac{2}{q-1}-1$ is
defined in Eq. (\ref{qk}) for $q\in[1,3),$ then
\begin{equation}
F_{q_k}^{-1} [\hat{f}(\xi)](x)= \int_{-\infty}^{\infty}
I_{(q_{k+1},{q_{k-1}})}[\hat{f}(\xi)] \otimes_{q_{k-1}}
e_{q_{k-1}}^{-i x \xi} d\xi,
\end{equation}
with the operator $I_{(q_{k+1},{q_{k-1}})}$ given in (\ref{i}). This
might constitute a first step for finding a representation of
$F_{q}^{-1} [\hat{f}(\xi)](x)$ for generic $\hat{f}(\xi) \in
L_1(R)$, which would be of great usefulness.

\end{document}